\documentclass[]{aa}

\def\rmit#1{{\it #1}}              %% italics (RR mode, Kluwer)
\def\specchar#1{{\sc #1}}
\def\FeI{\mbox{Fe\,\specchar{i}}}

\def\SiI{\mbox{Si\,\specchar{i}}}
\def\CaI{\mbox{Ca\,\specchar{i}}}
\def\HeI{\mbox{He\,\specchar{i}}}
\def\CaIIH{\mbox{Ca\,\specchar{ii}\,\,H}}       %% use \CaIIK\ for space

\def\eg{\rmit{e.g.}}

\def\arcsec{\hbox{$^{\prime\prime}$}}

\def\pun{\stackrel{}{\mbox{.}}}
\def\farcs{$\stackrel{\prime\prime}{\pun}$}

\usepackage{booktabs}
\usepackage{graphicx}
\usepackage{multirow}
\usepackage{color}
\usepackage[flushleft]{threeparttable}

\usepackage{array}
\newcolumntype{?}{@{\vrule width 2pt}}

\usepackage{hyperref}
\hypersetup{
    final=true,
    pageanchor=true,
    colorlinks=true,
    breaklinks=true,
    linkcolor=blue,
    citecolor=blue,
    urlcolor=blue,
    pdfpagemode=UseNone,
    pdftitle={},
    pdfauthor={T. Felipe et al.},
    pdfsubject={Solar Physics},
    pdfkeywords={Methods: observational -- Sun: photosphere -- Sun: chromosphere -- Sun: oscillations  -- sunspots -- Techniques: polarimetric}}

\titlerunning{Cutoff frequency in a sunspot umbra}   

\begin{document}

%\title{Flare ejected plasma produces changes in the photospheric magnetic field and heating events in a sunspot light-bridge}

%\title{How the time required to scan the spectropolarimetric profiles of chromospheric lines changes the inferred umbral flash atmospheres}

\title{Height variation of the cutoff frequency in a sunspot umbra}

\author{T. Felipe\inst{\ref{inst1},\ref{inst2}}
\and C. Kuckein\inst{\ref{inst3}}
\and I. Thaler\inst{\ref{inst4},\ref{inst5}}
}

%\author{Authors}
%\email{tobias@iac.es}

\institute{Instituto de Astrof\'{\i}sica de Canarias, 38205, C/ V\'{\i}a L{\'a}ctea, s/n, La Laguna, Tenerife, Spain\label{inst1}
\and 
Departamento de Astrof\'{\i}sica, Universidad de La Laguna, 38205, La Laguna, Tenerife, Spain\label{inst2} 
\and 
Leibniz-Institut f{\"u}r Astrophysik Potsdam (AIP), An der Sternwarte 16, 14482 Potsdam, Germany\label{inst3} 
\and
Kiepenheuer-Institut f\"ur Sonnenphysik, Sch\"oneckstr. 6, 79104 Freiburg, Germany\label{inst4}
\and
Racah Institute of Physics, The Hebrew University of Jerusalem, 91904 Jerusalem, Israel\label{inst5}
}

\abstract
{In the solar atmosphere, the acoustic cutoff frequency is a local quantity which depends on the atmospheric height. It separates the low-frequency evanescent waves from the high-frequency propagating waves.} 
{We measure the cutoff frequency of slow magnetoacoustic waves at various heights of a sunspot umbra and compare the results with the estimations from several analytical formulae.}
{We analyzed the oscillations in the umbra of a sunspot belonging to active region NOAA 12662 observed in the 10830 \AA\ spectral region with the GREGOR Infrared Spectrograph and in the \FeI\ 5435 \AA\ line with the GREGOR Fabry-P{\'e}rot Interferometer. Both instrumets are attached to the GREGOR telescope at the Observatorio del Teide, Tenerife, Spain. We have computed the phase and amplification spectra between the velocity measured from different pairs of lines that sample various heights of the solar atmosphere. The cutoff frequency and its height variation have been estimated from the inspection of the spectra.}
{At the deep umbral photosphere the cutoff frequency is around 5 mHz and it increases to 6 mHz at higher photospheric layers. At the chromosphere the cutoff is $\sim 3.1$ mHz. The comparison of the observationally determined cutoff with the theoretically predicted values reveals an agreement in the general trend and a reasonable match at the chromosphere, but also significant quantitative differences at the photosphere.}
{Our analyses show strong evidence of the variation of the cutoff frequency with height in a sunspot umbra, which is not fully accounted for by current analytical estimations. This result has implications for our understanding of wave propagation, the seismology of active regions, and the evaluation of heating mechanisms based on compressible waves.}


\maketitle

%%%%%%%%%%%%%%%%%%%%%%%%%%%%%%%%%%%%%%%%%%%%%%%%%%%%%%%%%%%%%%%%%%

\section{Introduction}

The pionering work by \citet{Lamb1909} showed that the cutoff frequency is a fundamental property of stratified mediums that determines the propagation of acoustic waves. Acoustic waves with frequencies above the cutoff can freely propagate in the atmosphere, whereas lower frequencies lead to non-propagating waves, also known as evanescent waves. Lamb's original work studied an isothermal atmosphere. In this case, the cutoff frequency is a global property and it is the same in the whole atmosphere. In the general case of a non-isothermal medium, such as the situation of the solar atmosphere, the cutoff frequency is a local quantity which depends on the atmospheric height.

Several works have attempted to derive an expression for the cutoff frequency in non-isothermal atmospheres \citep[\eg,][]{Schmitz+Fleck1998, Deubner+Gough1984, Musielak+etal2006}. However, different formulae for the cutoff are obtained depending on the selection of dependent and independent variables for the wave equations \citep{Schmitz+Fleck2003}. The situation described above does not take into account magnetic fields, which modify not only the existing temperature and pressure stratification, but also introduce new types of waves. The cutoffs of magnetohydrodynamic (MHD) waves in isothermal atmospheres with uniform magnetic field \citep[\eg,][]{Thomas1982, Thomas1983, Campos1986, Stark+Musielak1993, Roberts2006} and magnetic fluxtubes \citep[\eg,][]{Rae+Roberts1982, Roberts1983, Hammer+etal2010, Murawski+Musielak2010, Routh+etal2013, Perera+etal2015} have been thoroughly explored. Using the Solar Optical Telescope aboard Hinode spacecraft, \citet{Fujimura+Tsuneta2009} measured the phase differences between line-of-sight (LOS) velocity and magnetic flux in pores and intergranular magnetic structures, and concluded that the observations are consistent with standing sausage and/or kink MHD waves. 

This paper focuses on the analysis of oscillations in larger solar magnetic structures. In the presence of strong magnetic fields, $p-$modes travelling from the interior to the surface convert into slow and fast magnetoacoustic waves at the layer where the Alfv\'en speed and sound speed are similar \citep{Cally2006, Schunker+Cally2006, Khomenko+Collados2006, Cally2007}. The efficiency of the conversion to each of these modes depends on the angle between the wave vector and the magnetic field. In the region where the Alfv\'en speed is larger than the sound speed, the slow magnetoacoustic wave behaves like a sound wave, but it is guided along the field lines. This means that its velocity pertubations are along the field lines, while the wave can propagate in every direction except perpendicular to the magnetic field. Similar to the non-magnetic acoustic waves, in a gravitationally stratified medium they are affected by the cutoff. However, their cutoff value is reduced by a factor $\cos\theta$, where $\theta$ is the field inclination from the solar vertical, due to the reduced gravity along the magnetic field \citep{Bel+Leroy1977, McIntosh+Jefferies2006, Jefferies+etal2006}. This ramp effect allows the detection of low-frequency compressive waves in the chromosphere and corona of regions with strong magnetic fields \citep{Orrall1966,Giovanelli+etal1978a,Lites+etal1993,deMoortel+etal2002,dePontieu+etal2004,Vecchio+etal2007, Bloomfield+etal2007b}, and has been employed to reconstruct the magnetic field inclination in sunspots \citep{Tziotziou+etal2006,Yuan+etal2014,LohnerBottcher+etal2016}. An alternative process that leads to a reduction of the cutoff frequency is the radiative energy losses \citep{Roberts1983, Centeno+etal2006, Khomenko+etal2008b}, although the numerical simulations from \citet{Heggland+etal2011} indicate that field inclination is more relevant than radiation to explain long-period wave propagation.

In this work, we observationally quantify the acoustic cutoff frequency in a sunspot umbra. We focus on the analysis of slow magnetoacoustic waves propagating along vertical magnetic fields. The height variation of the cutoff is determined from multi-line observations. This paper is organized as follows. The observations and methods are described in Sect. \ref{sect:observations}. The analysis and results are presented in Sect. \ref{sect:wave_propagation} and discussed in Sect. \ref{sect:discussion}. Finally, conclusions are given in Sect. \ref{sect:conclusions}.

\begin{figure}[!ht] 
 \centering
 \includegraphics[width=9cm]{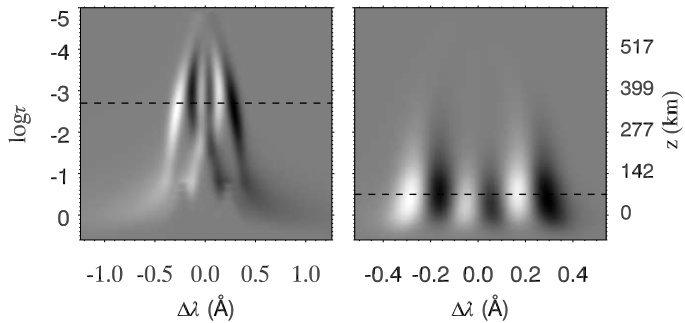}
  \caption{Response functions of the intensity to the velocity of the \SiI\ 10827 \AA\ (left panel) and \CaI\ 10839 \AA\ (right panel) lines in the average umbral atmosphere. Horizontal dashed lines indicate the optical depth selected for the analysis of the velocity from each line.}      
  \label{fig:RFs}
\end{figure}

\section{Sunspot observations and data analysis}
\label{sect:observations}

The sunspot in active region NOAA 12662 was observed with the 1.5-meter GREGOR solar telescope \citep{Schmidt+etal2012}
on 2017 June 17. The sunspot was located at the solar position $x=-387\arcsec$, $y=179\arcsec$ ($\mu=0.91$, with $\mu$ defined as the cosine of the heliocentric angle). Spectropolarimetric data was acquired with the GREGOR Infrared Spectrograph
\citep[GRIS;][]{Collados+etal2012} around the 1\,$\mu$m spectral window. The relevant lines
include the photospheric \ion{Si}{i} 10827\,\AA\ and \ion{Ca}{i} 10839\,\AA\ lines, together
with the chromospheric \ion{He}{i} 10830\,\AA\ triplet and two telluric lines. The later 
two lines were
used for the wavelength calibration on an absolute scale \citep{MartinezPillet+etal1997,Kuckein+etal2012b}. The slit (63\farcs5) was roughly fixed at the
center of the sunspot and covered the umbra, penumbra, and granulation on both sides.  
The full-Stokes spectra were taken with an exposure time of 100\,ms and 10 accumulations.
A total amount of 802 spectra were taken between 08:05 and 09:22\,UT.
The data reduction included dark and flat-field corrections as well as the standard 
polarimetric calibration \citep{Collados1999,Collados2003}. The data for the polarimetric calibration were obtained with the GREGOR polarimetric calibration unit \citep{Hofmann+etal2012}.

A second instrument, the GREGOR Fabry-P{\'e}rot Interferometer \citep[GFPI;][and references therein]{Puschmann+etal2012},
was simultaneously acquiring spectroscopic images of the sunspot in the photospheric non-splitting 
\ion{Fe}{i} 5435\,\AA\ line. Each scan along the line consisted of 20 wavelength positions 
with eight accumulations per position and an exposure time of 20 ms each. 
The cadence was about 23\,s in the $2\times2$ binned mode.
The image scale was 0\farcs081\,pixel$^{-1}$. The sunspot was placed in the center of 
the field-of-view (FOV) which had a size of 55\farcs7$\times$41\farcs5.
Observations with the GFPI started at 08:10 and stopped at 09:20\,UT. In this period of time, a total amount of 184 
scans were taken. 
The data reduction was accomplished using sTools \citep{Kuckein+etal2017}.
The pipeline carries out dark, flat-field, and prefilter curve corrections. The blueshift across the FOV was removed. 
The broad-band and narrow-band images were restored using Multi-Object Multi-Frame Blind Deconvolution 
\citep[MOMFBD;][]{Lofdahl2002,vanNoort+etal2005}. However, sporadic periods of poor seeing spoiled our long time series producing several incorrectly restored images. Therefore, to assure a continous analysis of the oscillations, we only focused on the non-restored data.

\begin{figure*}[!ht] 
 \centering
 \includegraphics[width=18cm]{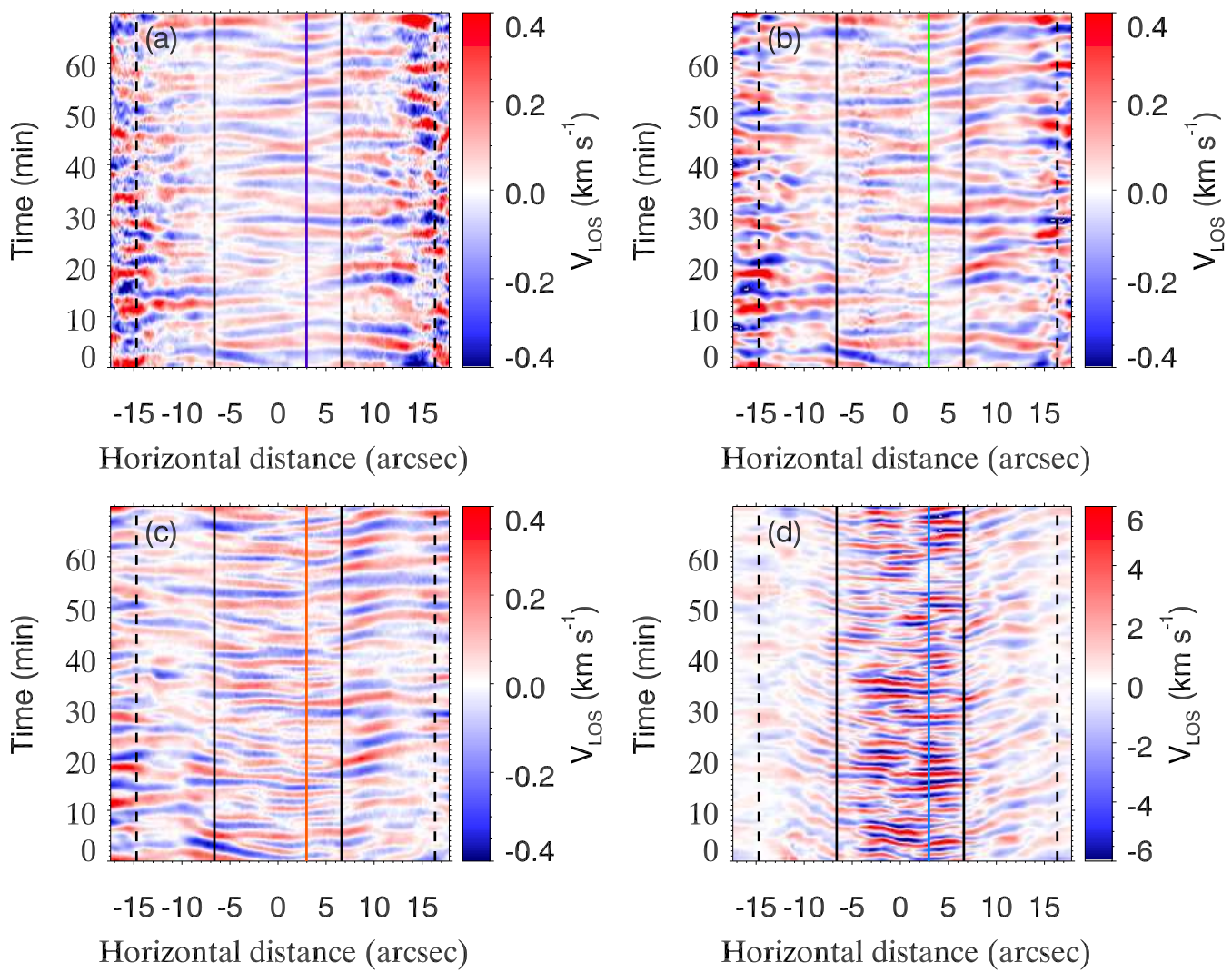}
  \caption{LOS velocity measured with the \CaI\ 10839 \AA\ line (a), the \SiI\ 10827 \AA\ line (b), the  \FeI\ 5435 \AA\ line (c), and \HeI\ 10830 \AA\ triplet (d). The signals have been filtered in the frequency range between 1 and 15 mHz. Positive velocities (red) correspond to downflows and negative velocities (blue) to upflows. The vertical colored lines mark the umbral location illustrated in Fig. \ref{fig:velocity_x}}      
  \label{fig:velocity_maps}
\end{figure*}

\begin{figure*}[!ht] 
 \centering
 \includegraphics[width=18cm]{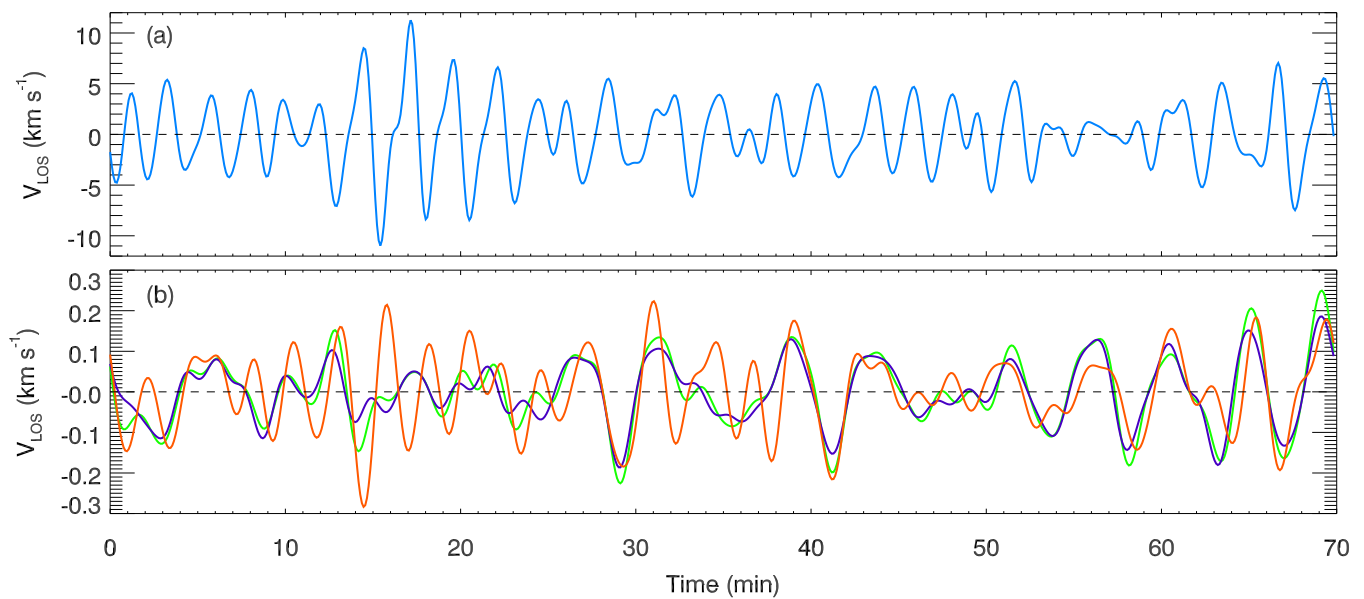}
  \caption{Top panel: Temporal evolution of the LOS velocity at a selected umbral location measured with the \HeI\ 10830 \AA\ triplet; bottom panel: Temporal evolution of the LOS velocity at the same umbral location measured with the \CaI\ 10839 \AA\ (violet), \SiI\ 10827 \AA\ (green), and \FeI\ 5435 \AA\ (orange) lines. The umbral location is indicated by the vertical colored lines in Fig. \ref{fig:velocity_maps}}      
  \label{fig:velocity_x}
\end{figure*}

For the photospheric lines (\SiI\ 10827 \AA\ and \CaI\ 10839 \AA), the temporal evolution of the sunspot atmosphere has been determined using the SIR code \citep{RuizCobo+delToroIniesta1992}. For each spatial position and time step, SIR provides the vertical stratification of the magnetic and thermodynamic atmospheric parameters retrieved from a process in which an initial guess atmosphere \citep[in our case, the cool sunspot model from][]{Collados+etal1994} is iteratively modified until it reproduces the observed Stokes parameters. We have performed independent single-component inversions for the \SiI\ 10827 \AA\ and \CaI\ 10839 \AA\ lines. In both cases, the temperature stratification is modified with three nodes, while the other physical parameters (LOS velocity, microturbulence, field strength, and inclination and azimuth angles of the magnetic field) are changed with two nodes. In this paper, we focus on the analysis of the LOS velocities in the sunspot umbra. We have selected the optical depth where the response of the lines to the velocity is maximum (see Fig. \ref{fig:RFs}). In the case of the \CaI\ 10839 \AA\ line, it corresponds to $\log\tau\sim-0.5$, while for the \SiI\ 10827 \AA\ it is $\log\tau\sim-2.7$. In the following, we will refer to those optical depths when we discuss \CaI\ 10839 \AA\ and \SiI\ 10827 \AA\ LOS velocities. The response functions illustrated in Fig. \ref{fig:RFs} were computed for the model obtained as the average of atmospheres retrieved from the inversion of the \SiI\ 10827 \AA\ line in the umbra. In this model, the maximum response of the \CaI\ 10839 \AA\ and \SiI\ 10827 \AA\ lines to the velocity corresponds to a geometrical height of 64 and 340 km, respectively, with $z=0$ located at the optical depth where $\log\tau=0$. Top panels of Fig. \ref{fig:velocity_maps} illustrate the velocity maps of these two lines. The LOS velocity from the chromospheric \HeI\ 10830 \AA\ line (Fig. \ref{fig:velocity_maps}d) has been retrieved from a Milne-Eddington inversion using the code MELANIE \citep{SocasNavarro2001}. Figure \ref{fig:velocity_x} shows the temporal evolution of the LOS velocities at a selected umbral location where the strongest \HeI\ 10830 \AA\ velocity oscillations are detected. Around $t=16$ min its peak-to-peak amplitude is 22 km/s, which is significatively larger than the amplitude of any previously detected umbral oscillations in \HeI\ 10830 \AA\ \citep[generally below 14 km/s, \eg,][]{Centeno+etal2006, Felipe+etal2010b}.

\begin{figure}[!ht] 
 \centering
 \includegraphics[width=9cm]{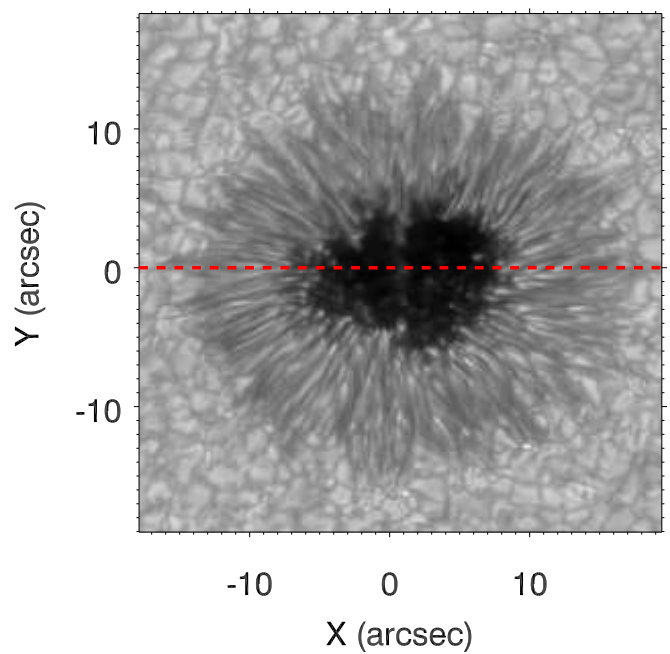}
  \caption{Reconstructed broad-band image of the sunspot NOAA 12662 on June 17 2017 at 08:12:45 UT from GFPI. The red dashed line indicates the location of the GRIS slit determined following the method described in Sect. \ref{sect:observations}. }      
  \label{fig:FOV_GFPI}
\end{figure}

The \FeI\ 5435 \AA\ line observed with GFPI is formed at high photospheric layers. Using NLTE modeling of the spectral line radiative transfer and numerical simulations of granular convection, \citet{BelloGonzalez+etal2010a} estimated a formation height of 500 km in granules and 620 km in intergranules. Assuming that in the sunspot umbra the response function of the line to the velocity peaks at the same density as in intergranules, the formation height of the \FeI\ 5435 \AA\ line in our average umbra is 510 km. The LOS velocity of the \FeI\ 5435 \AA\ line has been determined by measuring the Doppler shift using a second order polynomial fit to the intensity core of the line. The \FeI\ 5435 \AA\ line is insensitive to the magnetic field (Land\'e factor $g=0$) and, thus, the estimation of the core of the line is not affected by Zeeman splitting. Note that GFPI data provides the temporal evolution in a bidimensional surface. The following approach has been carried out in order to construct GFPI temporal series co-spatial to the location of the GRIS slit. First, \FeI\ 5435 \AA\ velocities from GFPI have been converted to a new grid with the spatial size and the time steps of the GRIS data. Only the temporal span with simultaneous data acquisition in both instruments (between 08:10 and 09:20 UT) has been used. Second, the GRIS \CaI\ 10839 \AA\ and the GFPI \FeI\ 5435 \AA\ velocities have been filtered in the frequency band between 1.5 and 3 mHz. This way, we eliminate the propagating high frequency waves (above the expected cutoff frequency) and keep only the evanescent waves. Third, assuming that the GRIS slit is aligned with the X axis of the GFPI field-of-view (FOV), we have iteratively computed the cross-correlation between the filtered \CaI\ 10839 \AA\ waves and the filtered \FeI\ 5435 \AA\ waves at different positions in the Y direction and applying different shifts in the X direction. Since those waves are evanescent, they oscillate in phase for different heights. We have determined the location of the GRIS slit in the GFPI FOV as the Y coordinate and the X shift that shows the highest correlation. The position of the slit on a region of the GFPI FOV is illustrated in Fig. \ref{fig:FOV_GFPI}. Finally, that region has been extracted from the remapped temporal series obtained at the first step. Figure \ref{fig:velocity_maps}c shows the temporal evolution of the \FeI\ 5435 \AA\ velocity at the location of the GRIS slit.

The data from 2017 June 17 have been complemented with another data set from 2007 August 28 obtained with the Vacuum Tower Telescope at the Observatorio del Teide. These data correspond to the active region NOAA 10969, which was located near disk center ($\mu=0.96$), and the spectra were obtained from a slit placed over the center of a sunspot. They were described and analyzed in \citet{Felipe+etal2010b}. In this work, we will use Doppler velocities of the \CaIIH\ and \HeI\ 10830 \AA\ lines in order to include the analysis of two chromospheric lines obtained simultaneously.

\begin{figure}[!ht] 
 \centering
 \includegraphics[width=9cm]{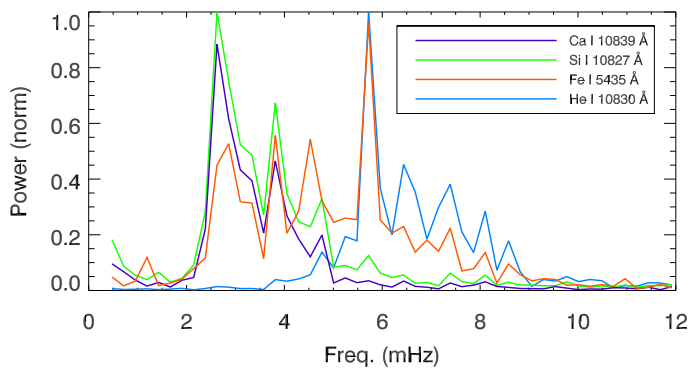}
  \caption{Average umbral power spectra of the LOS velocities obtained with the \CaI\ 10839 \AA\ (violet), \SiI\ 10827 \AA\ (green), \FeI\ 5435 \AA\ (orange), and \HeI\ 10830 \AA\ (blue) lines. The power of the three former lines is normalized to the maximum power of the \SiI\ 10827 \AA\ line, while the \HeI\ 10830 \AA\ is normalized to its maximum power.}      
  \label{fig:power}
\end{figure}

\section{Results}
\label{sect:wave_propagation}

\subsection{Power spectra}

Figure \ref{fig:power} shows the umbral power (averaged for all the umbra, which is delimited by the black solid vertical lines in Fig. \ref{fig:velocity_maps}) of the LOS velocity of the four spectral lines from the 2017 June 17 dataset. As expected, most of the power of the photospheric \SiI\ 10827 \AA\ (green) and \CaI\ 10839 \AA\ (violet) lines is concentrated in the 5-minute band, in the frequency band between 2 and 4 mHz. At the higher formation height of the \FeI\ 5435 \AA\ line (orange) the power in the 5-minute band is still significant, but the strongest peak is found at 5.7 mHz. At the chromosphere (\HeI\ 10830 \AA, blue line) the power spectra shows a broad distribution, with several power peaks in the frequency range between 5 and 9 mHz (3-minute band). The maximum power is located also at 5.7 mHz. 

Interestingly, all the oscillatory signals, from the deep photosphere (\CaI\ 10839 \AA) to the chromosphere (\HeI\ 10830 \AA), exhibit power peaks at some shared frequencies. However, the relative power between them depends on the formation height of the line. The main chromospheric peaks (5.7, 6.4, 7.4, 8.1, and 8.6 mHz) are already present in the velocity oscillations measured with the \CaI\ 10839 \AA\ and \SiI\ 10827 \AA\ lines, although their photospheric power is insignificant in comparison with that in the 5-minute band. This finding supports the results from \citet{Centeno+etal2006}, who proposed that the chromospheric 3-minute power reaches the higher atmospheric layers through wave propagation from the photosphere. The amplitude of the propagating high-frequency waves (above the cutoff frequency) increases more than that of the non-propagating (evanescent) low-frequency waves. This way, waves in the 3-minute band dominate at chromospheric heights.
Taking the former into account, an approximate value of the cutoff frequency can be estimated from an examination of the height variation of the power. For the \CaI\ 10839 \AA\ line (deep photosphere), the power at 5.2 mHz is larger than the power at any higher frequency. Althought it increases with height, the increase rate of higher frequencies is larger. For example, for the \SiI\ 10827 \AA\ line the power at the peak located 5.7 mHz is already larger than the power peak at 5.2 mHz, and 5.7 mHz oscillations dominate the high photosphere (\FeI\ 5435 \AA) and chromosphere (\HeI\ 10830 \AA). This fact indicates that the cutoff frequency of the umbral atmosphere is around 5.5 mHz. However, the cutoff is not a constant quantity, since it changes with the atmospheric stratification. In the following section we explore the variation of the local cutoff values.

\begin{figure}[!ht] 
 \centering
 \includegraphics[width=9cm]{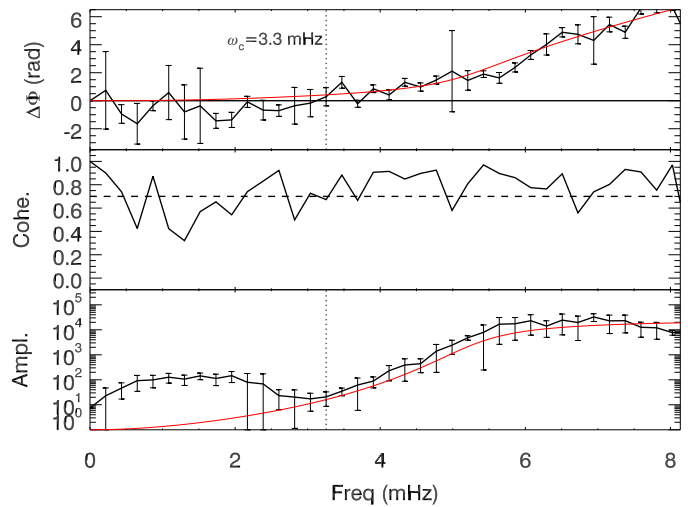}
  \caption{Phase (top panel), coherence (middle panel), and amplification (bottom panel) spectra between the LOS velocities of the \CaI\ 10839 \AA\ and the \HeI\ 10830 \AA. Black line indicates the measured values and the red lines correspond to the best fit to a theoretical model. Errors are determined as described in the text.}
      
  \label{fig:dphase_CaHe}
\end{figure}

\begin{figure}[!ht] 
 \centering
 \includegraphics[width=9cm]{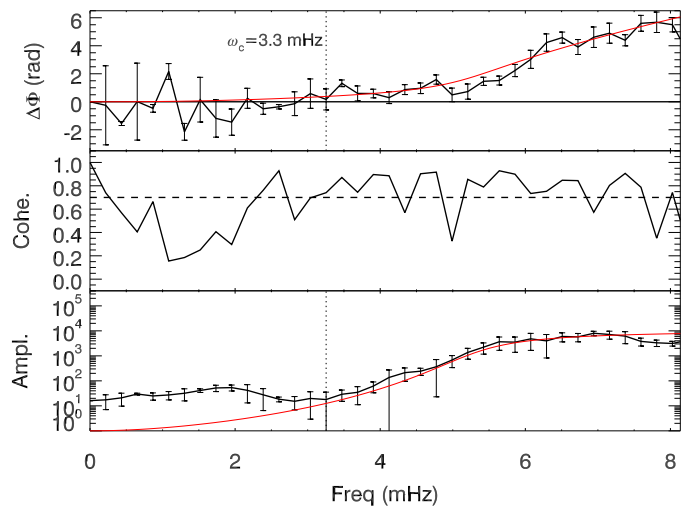}
  \caption{Same as Fig. \ref{fig:dphase_CaHe} but for the line pair \SiI\ 10827 \AA\ and \HeI\ 10830 \AA.}
      
  \label{fig:dphase_SiHe}
\end{figure}

\subsection{Phase difference and cutoff}
\label{Sect:cutoff}
The phase difference ($\Delta\phi$) as a function of frequency between velocity signals at two different heights has been computed for several pairs of lines. As a first example, Fig. \ref{fig:dphase_CaHe} illustrates the phase difference between the two lines with larger difference in formation height, the \CaI\ 10839 \AA\ and \HeI\ 10830 \AA\ lines. The phase difference at each spatial position is obtained by subtracting the phase of the former line to the phase of the later. Thus, a positive (negative) value indicates upwards (downwards) wave propagation. The $2\pi$ indetermination in the phase difference has been manually unwrapped by adding $2\pi$ to the $\Delta\phi$ negative values at high frequencies, in order to maintain the smooth trend of $\Delta\phi$ measured at lower frequencies. The black solid line in the top panel of Fig. \ref{fig:dphase_CaHe} shows the average of the phase difference inside the umbra. For each frequency, an histogram with the relative occurrence of a given value of $\Delta\phi$ (in 0.3 rad bins) over all pixels in the umbra with a magnetic field inclination below 20$^{\circ}$ (in the local solar frame, according to the inversions of the \SiI\ 10827 \AA\ line) has been computed. Only the spatial points whose $\Delta\phi$ is in a bin with more than 7\% occurrence are included in the average. This way, we avoid contamination from outlayers. The bars in the phase difference correspond to the standard deviation of the averaged points. 

Following \citet{Centeno+etal2006}, we have also computed the coherence and amplification spectra. The coherence depends on the frequency and takes values in the range [0,1]. A higher coherence indicates that the phase difference between two signals is characteristic of their oscillations. We consider the phase difference to be significant for coherence values above 0.7. The amplification spectra is defined as the ratio between the power of the two signals, averaged in the umbra. The bottom panel of Fig. \ref{fig:dphase_CaHe} (and the other figures with the same format) depicts the standard deviation of the amplification for the umbral points.

The phase and amplification spectra have been fitted to a theoretical model of slow mode wave propagation in a gravitationally stratified isothermal atmosphere that allows radiative losses \citep[see][for details]{Centeno+etal2006, Felipe+etal2010b}. The fitting is performed manually by tunning three free parameters: atmospheric temperature ($T$), height difference between the formation height of the two lines ($\Delta z$), and the characteristic timescale of the radiative losses ($\tau_{\rm R}$). With the inclusion of the radiative losses in the modeling, there is not a strict cutoff frequency which separates the propagating (high-frequency) waves from the non-propagating (evanescent, low-frequency) waves. Instead, all frequencies are propagated and reflected simultaneously, and the ratio of propagation versus reflection grows with frequency. Although there is no cutoff frequency, an effective cutoff frequency, separating mainly propagating waves from mainly non-propagating waves, can be defined. The parameters determined from this fitting for all the analyzed pairs of lines are displayed in Table \ref{table:parameters}.

\begin{figure}[!ht] 
 \centering
 \includegraphics[width=9cm]{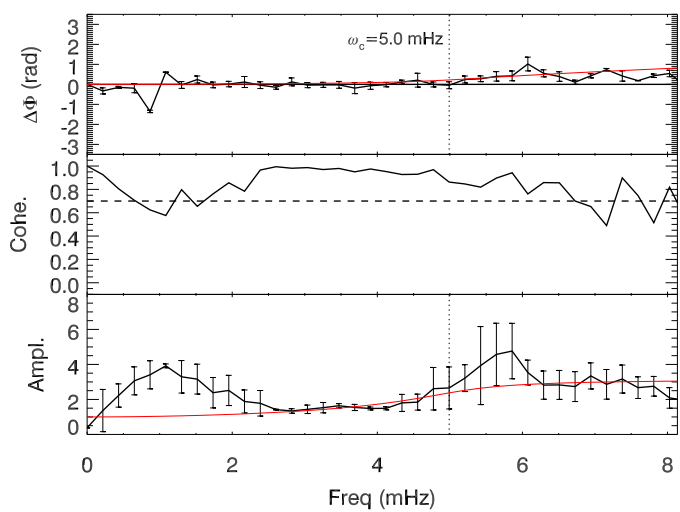}
  \caption{Same as Fig. \ref{fig:dphase_CaHe} but for the line pair \CaI\ 10839 \AA\ and \SiI\ 10827 \AA.}
      
  \label{fig:dphase_CaSi}
\end{figure}

\begin{figure}[!ht] 
 \centering
 \includegraphics[width=9cm]{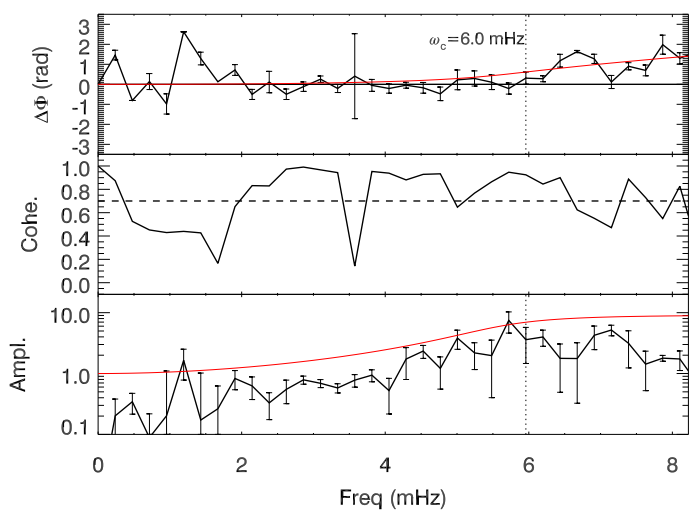}
  \caption{Same as Fig. \ref{fig:dphase_CaHe} but for the line pair \SiI\ 10827 \AA\ and \FeI\ 5434.534 \AA.}
      
  \label{fig:dphase_SiFe}
\end{figure}

\begin{table}
\begin{center}
\caption[]{\label{table:parameters}
          {Parameters for phase and amplification spectra fits and observed cutoff frequency}}
\begin{tabular}{ccccc}
\hline\noalign{\smallskip}
Line Pair			& 	T (K)	& $\Delta z$ (km)	&	$\tau_{\rm R}$ (s)	& $\omega_{\rm c}$ (mHz)	\\
\hline\noalign{\smallskip}	
\CaI\ - \SiI\ 		& 	4000	& 130			& 30				&	5.0		\\
\SiI\ - \FeI\ 		& 	3000	& 200			& 20				&	6.0		\\
\FeI\ - \HeI\ 		& 	3500	& 750 			& 40				&	3.6		\\
\CaIIH\ - \HeI\  		& 	7000	& 150			& 20				&	3.1		\\
\CaI\ - \HeI\ 	& 	4000	& 1100			& 40				&	3.3		\\
\SiI\ - \HeI\ 	& 	4000	& 1000			& 40				&	3.3		\\
\hline

\end{tabular}

\begin{tablenotes}
\small
 \item {The first column indicates the pair of lines used in the analysis, second to fourth columns are the temperature ($T$), difference in formation heights ($\Delta z$), and characteristic timescale of the radiative losses ($\tau_{\rm R}$) retrieved from the fit of the phase and amplification spectra to a model of linear wave propagation in a gravitationally stratifed atmosphere with radiative losses, and the last column in the observationally measured cutoff frequency ($\omega_{\rm c}$).}
\end{tablenotes}

\end{center}
\end{table}

Figure \ref{fig:dphase_CaHe} indicates that propagating waves go upwards in the umbral atmosphere. The examination of the phase and amplification spectra allow us to estimate the cutoff frequency. For frequencies with $\Delta\phi$ around $0^{\circ}$ and a low amplitude increase, the observed oscillations correspond to evanescent waves. On the contrary, frequencies with $\Delta\phi>0^{\circ}$ and a larger amplitude increase are indicative of upward propagating waves. However, the solar atmosphere is a stratified and inhomogenous medium and, thus, the cutoff frequency is a local quantity that depends on the atmospheric height. We have probed the height dependence of the cutoff frequency by estimating it for several pairs of lines. We defined the observed cutoff frequency $\omega_{\rm c}$ as the highest frequency whose phase difference is not undoubtedly different from zero. In order to ensure this condition, for each pair of lines we selected the highest frequency with $\Delta\phi$ and its standard deviation out of a value of zero and with a coherence higher than the chosen confidence threshold of 0.7. In addition, we verified that the  increasing trend of the phase difference is maintained for the following frequency values and that this increasing phase difference is accompanied by an increase in the amplification spectra.

\begin{figure}[!ht] 
 \centering
 \includegraphics[width=9cm]{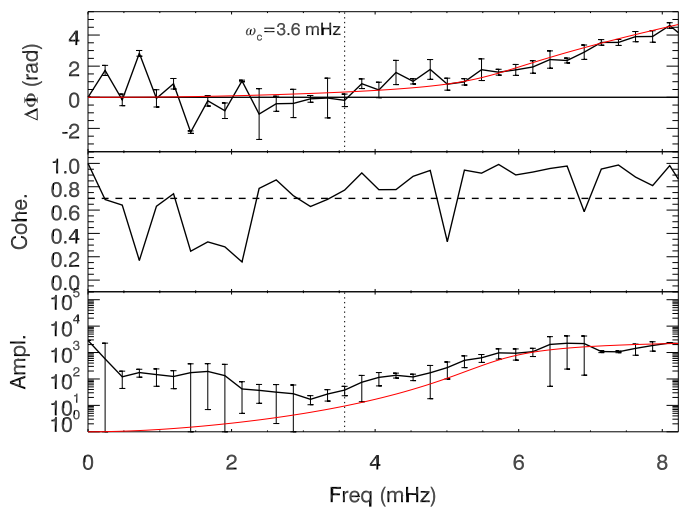}
  \caption{Same as Fig. \ref{fig:dphase_CaHe} but for the line pair \FeI\ 5434.534 \AA\ and \HeI\ 10830 \AA.}
      
  \label{fig:dphase_FeHe}
\end{figure}

\begin{figure}[!ht] 
 \centering
 \includegraphics[width=9cm]{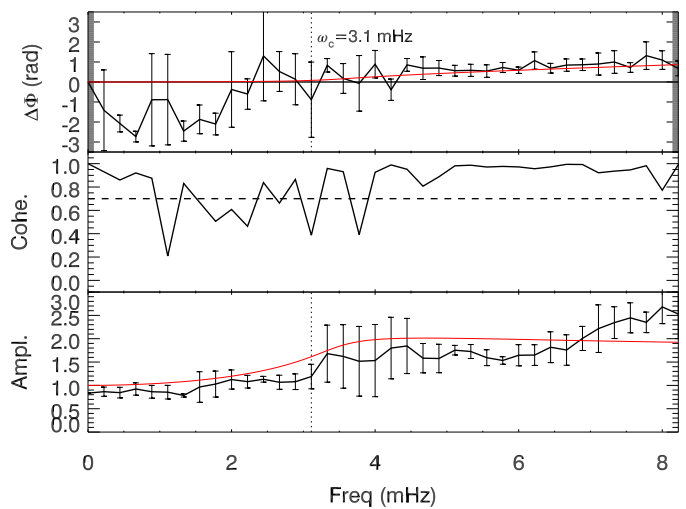}
  \caption{Same as Fig. \ref{fig:dphase_CaHe} but for the line pair \CaIIH\ and \HeI\ 10830 \AA.}
      
  \label{fig:dphase_CaIIHHe}
\end{figure}

The observationally estimated cutoff frequencies are shown in the last column of Table \ref{table:parameters}. They have been determined from the measurements illustrated in Figs. \ref{fig:dphase_CaHe}-\ref{fig:dphase_CaIIHHe}. The analysis of pairs of photospheric and chromospheric lines (\CaI\ 10839 \AA\ - \HeI\ 10830 \AA, \SiI\ 10827 \AA\ - \HeI\ 10830 \AA, Figs. \ref{fig:dphase_CaHe} and \ref{fig:dphase_SiHe}, respectively) shows a similar behavior. For low frequencies, the phase difference is around zero, with noisy measurements and low coherence. For frequencies above 3.3 mHz, the phase and amplification spectra start to progresively increase, indicating upward wave propagation. This value was chosen as the cutoff frequency, since it is evident that waves with higher frequencies have propagated at least at some heights of the umbral atmosphere. Note that this cutoff value must be interpreted as the minimum value of the cutoff between the formation height of the photospheric and chromospheric lines. Even though between these two layers waves can become evanescent in some regions, we can detect the phase difference produced in the heights where waves were able to propagate. We have sampled the atmospheric variation of the cutoff value by exploring pairs of lines with smaller differences in formation height. Thus, these pairs of lines sample the cutoff at localized layers of the atmosphere.    

Figure \ref{fig:dphase_CaSi} illustrates the results for the \CaI\ 10839 \AA\ - \SiI\ 10827 \AA\ pair, that is, for wave propagation between the deep photosphere and medium photosphere. Due to the small height difference between the formation height of these two lines, the values of $\Delta\phi$ are low, below 1 rad. Phase differences different from zero are detected for frequencies higher than 5.0 mHz. This cutoff value must be interpreted as an upper limit of the actual cutoff. Waves with lower frequencies could propagate, but their low phase difference (due to the small difference in formation height between both lines) might be undetectable for the uncertanties of our measurements. The amplification spectra seems to point to that, since an increase in the amplification is found for frequencies above 4.1 mHz, although this trend is also masked by the errors of the amplification spectra.   

Waves between the photosphere (\SiI\ 10827 \AA) and high photosphere (\FeI\ 5435 \AA) are analyzed in Fig. \ref{fig:dphase_SiFe}. The phase difference spectra exhibits a cutoff frequency of 6.0 mHz although, similarly to the previous case, the amplification spectra points to wave propagation for lower frequencies. Interestingly, the undoubtedly evanescent waves (frequencies below 4 mHz) show lower amplitude at the formation height of the \FeI\ 5435 \AA\ line than at the \SiI\ 10827 \AA\ line. This is consistent with the power spectra illustrated in Fig. \ref{fig:power}, but cannot be understood from a wave model (see red line in the bottom panel of Fig. \ref{fig:dphase_SiFe}). We consider that this reduction in the velocity amplitude of the \FeI\ 5435 \AA\ line is due to the different acquisition strategy, since the data from the Fabry-P\'erot interferometer has lower spectral resolution than that from the slit spectrograph and it requires more time to scan the spectral profile of the line.

Figure \ref{fig:dphase_FeHe} shows that waves with frequency above 3.6 mHz can travel from the high photosphere (\FeI\ 5435 \AA) to the chromosphere (\HeI\ 10830 \AA). In this case, the estimation of the cutoff frequency from the phase difference spectra is in agreement with that from the amplification spectra. Finally, the phase difference between the chromospheric \CaIIH\ and \HeI\ 10830 \AA lines (Fig. \ref{fig:dphase_CaIIHHe}) shows continuous positive phase shifts for frequencies above 4.2 mHz, while lower frequencies (3.3 and 4 mHz) also exhibit a reliable positive phase difference. Based on these data and the variation found for the amplification spectra, we determine a cutoff frequecy of 3.1 mHz.

\section{Discussion}
\label{sect:discussion}

The cutoff frequency is a well-known property of gravitationally stratified atmospheres. Despite its strong impact on wave propagation, there is not a clear definition of its value, since the analytical expression obtained for its computation depends on the choice of the dependent and independent variables used in the wave equation \citep{Schmitz+Fleck2003}. Only in the case of an isothermal atmosphere, a correct formula for the cutoff can be determined as \citep{Lamb1909}:

\begin{equation}
\omega_{\rm C1}=\frac{c_{\rm s}}{2H_{\rm p}},
\label{eq:wc1}
\end{equation}

\noindent where $c_{\rm s}$ is the sound speed and $H_{\rm p}$ is the pressure scale height. However, the solar atmosphere is not isothermal, and this expression is not generaly applicable. The most commonly form used by helioseismologists is

\begin{equation}
\omega_{\rm C2}=\frac{c_{\rm s}}{2H_{\rm \rho}}\Big (1-2\frac{dH_{\rm \rho}}{dz}\Big )^{1/2},
\label{eq:wc2}
\end{equation}

\noindent where $H_{\rm \rho}$ is the density scale height. This expression produces a sharp spike just below the solar surface, which is inconsistent with the WKB approximation that is usually employed for the study of linear acoustic waves. This approximation is only valid for waves with wavelengths shorter than the characteristic scales of the atmosphere. Some studies using the WKB approximation use the isothermal expression (Eq. \ref{eq:wc1}), even for non-isothermal problems \citep{Cally2007, Moradi+Cally2008}. The WKB approximation has also been used to derive a local dispersion relation and cutoff frequency for MHD waves in an isothermal solar atmosphere \citep{Thomas1982, Thomas1983, Campos1986}. In addition, in models with radiative losses there is not a clear cutoff frequency, but a pseudo-cutoff can be defined \citep{Centeno+etal2006}. This effective cutoff can be significantly lower than that obtained for the adiabatic case. 

Recently, \citet{Wisniewska+etal2016} have estimated the cutoff frequency of the solar quiet-Sun atmosphere (including its variation with the atmospheric height) and compared the results with the formulae derived from different theories. In this paper we aim to perform a similar estimation for a sunspot umbra atmosphere. In addition to Eqs. \ref{eq:wc1} (assuming that it changes locally) and \ref{eq:wc2}, we have also evaluated the expression derived by \citet{Schmitz+Fleck1998}

\begin{equation}
\omega_{\rm C3}=\omega_{\rm C1}\Big (1-2\frac{\frac{dc_{\rm s}}{dz}}{\omega_{\rm C1}}\Big )^{1/2}.
\label{eq:wc3}
\end{equation} 

\noindent The above formulae were derived for acoustic waves, and do not take into account the presence of the magnetic field. However, we expect them to be applicable to the slow magnetoacoustic waves analyzed in this work, since they have an acoustic-like behavior (we are observing an umbral atmosphere region where the Alfv\'en speed $v_{\rm A}$ is higher than the sound speed $c_{\rm s}$). We have also explored a specific formula for cutoff frequencies obtained for magnetoacoustic waves in an isothermal solar atmosphere permeated by an uniform vertical magnetic field. This expression derived by \citet{Roberts2006} is given by

\begin{equation}
\omega_{\rm C4}=c_{\rm t}\Big [\frac{1}{4H_{\rm p}^2} \Big (\frac{c_{\rm t}}{c_{\rm s}}\Big )^4-\frac{1}{2}\gamma g\frac{\partial}{\partial z}\Big (\frac{c_{\rm t}^2}{c_{\rm s}^4}\Big )+\frac{1}{v_{\rm A}^2}\Big (N^2+\frac{g}{H_{\rm p}}\frac{c_{\rm t}^2}{c_{\rm s}^2}\Big )\Big ]^{1/2} 
\label{eq:wc4}
\end{equation}

\noindent where $N^2$ is the squared Brunt-V\"ais\"al\"a frequency, $c_{\rm t}=c_{\rm s}v_{\rm A}/\sqrt{c_{\rm s}^2+v_{\rm A}^2}$ is the cusp speed, $\gamma$ is the adiabatic index, and $g$ is the gravity.

Our approach of measuring the cutoff is slightly different from that performed by \citet{Wisniewska+etal2016}. First, we have estimated the phase shift between different combinations of pairs of lines, instead of only between the line with highest formation height and the rest. This way, we can probe localized layers of the atmosphere and restrict the derived cutoff to a more constrained height. Second, in addition to the analysis of the phase difference, we have complemented the determination of the cutoff with the amplification spectra.

\begin{figure}[!ht] 
 \centering
 \includegraphics[width=9cm]{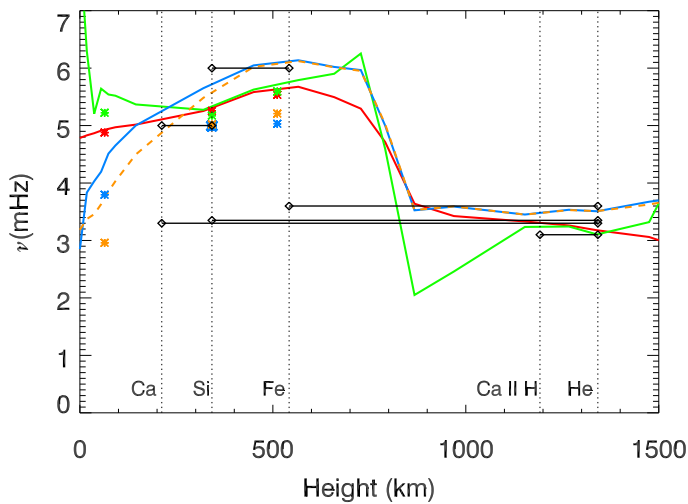}
  \caption{Variation of the cutoff frequency in the umbral atmosphere. Color lines represent the cutoff values computed for \citet{Avrett1981} model and the color asterisks correspond to the cutoffs calculated for the inverted umbral atmosphere. The color indicates the formula used: red (Eq. \ref{eq:wc1}), green (Eq. \ref{eq:wc2}), blue (Eq. \ref{eq:wc3}), and orange (Eq. \ref{eq:wc4}). Horizontal black lines with diamonds indicate the observational cutoffs as determined from Figs. \ref{fig:dphase_CaHe}-\ref{fig:dphase_CaIIHHe} (see Table \ref{table:parameters}). The black lines cover the range of heights between the formation height of the two lines used for each measurement. Vertical dotted lines mark the formation height of the spectral lines as determined from the analysis of wave propagation.}      
  \label{fig:cutoffs}
\end{figure}

The results are illustrated in Fig. \ref{fig:cutoffs}. We have computed the analytical cutoff using Eqs. \ref{eq:wc1}, \ref{eq:wc2}, \ref{eq:wc3}, or \ref{eq:wc4} for the atmospheric values retrieved from the inversions of the \CaI\ 10839 \AA\ and \SiI\ 10827 \AA\ lines. They are represented by red, green, blue, and orange asterisks, respectively. The height of these data is chosen at the maximum of the response function of the \CaI\ 10839 \AA, \SiI\ 10827 \AA, and \FeI\ 5435 \AA\ lines. The analytical cutoffs of the former were obtained from the inversions of the \CaI\ 10839 \AA\ line, while the other two were computed from the stratification retrieved from the \SiI\ 10827 \AA\ inversion. Note that the sensitivity of this inversion to the formation height of the \FeI\ 5435 \AA\ line is low. 

The observational cutoffs from Sect. \ref{Sect:cutoff} are indicated by the black horizontal lines. They link the two atmospheric heights (indicated by diamonds) used for computing the corresponding phase and amplification spectra. In these cases, the formation heights are determined based on the analysis of the wave propagation. The formation height of the \SiI\ 10827 \AA\ line has been set to the height given by the response function (same height from the previous paragraph), while the rest of the formation heights are derived from the height differences indicated in Table \ref{table:parameters}. The difference in formation height between the \SiI\ 10827 \AA\ and \FeI\ 5435 \AA\ lines obtained from the analysis of the oscillations is consistent with that retrieved from the evaluation of their response functions. On the contrary, the peak of the response function of the \CaI\ 10839 \AA\ line is deeper than the location expected from the phase shift of its oscillations with respect to those from the \SiI\ 10827 \AA\ line.

By plotting the measured cutoff over a range of heights (black horizontal lines), we explicitly indicate that the atmospheric layer with that value as local cutoff frequency is undetermined. The observational cutoffs must be interpreted as the minimum value of the actual cutoff between the two heights used in each measurement. However, thanks to the use of pairs of lines with close formation heights and the combination of many spectral lines, some conclusions about the stratification of the cutoff frequency can be extracted: (i) at the deep photosphere, its value is around 5 mHz, as shown by the analysis of the line pair \CaI\ 10839 \AA\ - \SiI\ 10827 \AA; (ii) the cutoff increases at higher photospheric layers, since the line pair \SiI\ 10827 \AA\ - \FeI\ 5435 \AA\ exhibits a value around 6 mHz; and (iii) at the chromosphere the cutoff is significantly lower, around 3.1 mHz. The later is proven by the analysis of the pair of chromospheric lines \CaIIH\ - \HeI\ 10830 \AA\ and by the cutoff estimated from the \CaI\ 10839 \AA\ - \HeI\ 10830 \AA\ and \SiI\ 10827 \AA\ - \HeI\ 10830 \AA\ pairs: since we concluded that from the deep photosphere to the high photosphere the cutoff is higher than 5 mHz, the measured phase difference for frequencies around 3.3 mHz must be produced at higher layers. 

Following \citet{Wisniewska+etal2016}, we have compared the observational results of the cutoff with the values computed using different formulae from the literature in a standard model \citep[in our case, the umbral atmosphere from][]{Avrett1981}. The four calculated cutoffs show some common properties that are confirmed by our observational measurements. The cutoff increases from the base of the photosphere to higher layers (up to $\sim 600-700$ km, depending on the formula), while at the chromosphere it is much lower, slightly above 3 mHz. \citet{Roberts2006} cutoff expression (Eq. \ref{eq:wc4}) for magnetoacoustic waves is almost indistinguishable from \citet{Schmitz+Fleck1998} formula (Eq. \ref{eq:wc3}) from the high photosphere to higher layers. The chromospheric cutoffs $\omega_{\rm C1}$ and $\omega_{\rm C2}$ show a remarkable agreement with the observational values, assuming that \citet{Avrett1981} model describes the umbra of this sunspot. At the photosphere, all the cutoff values calculated from \citet{Avrett1981} model and from the inversions (asterisks, using the same analytical expressions) exhibit some discrepancies with the observations. The cutoff retrieved from the inversions using \citet{Roberts2006} formula (orange asterisks) shows a perfect match at the formation height of the \SiI\ 10827 \AA\ line, although it does not reproduce the higher value observed at the high photosphere, from the analysis of the \SiI\ 10827 \AA\ and \FeI\ 5435 \AA\ pair of lines.

\begin{figure}[!ht] 
 \centering
 \includegraphics[width=9cm]{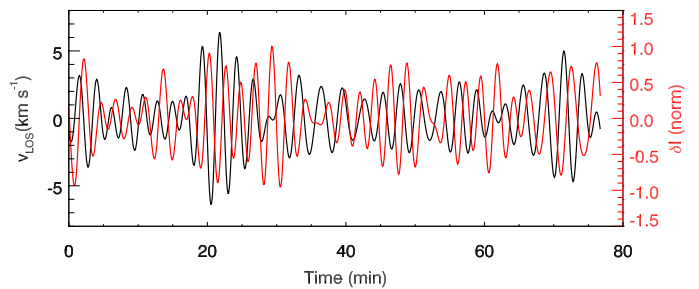}
  \caption{Temporal evolution of the LOS velocity (black line) and core intensity (red line) of the \HeI\ 10830 \AA\ filtered in the 4.5-8 mHz frequency range and spatially averaged for all the umbral spatial locations.}      
  \label{fig:phase_V_I}
\end{figure}

The presence of cutoff frequencies implies that waves can be trapped in the solar atmosphere due to wave reflections, and a cavity can potentially be formed \citep{Schmitz+Fleck1992}. The phase relations between the velocity and intensity can be used to determine whether the oscillations are propagating waves or standing waves \citep[\eg,][]{Fujimura+Tsuneta2009}. In an adiabatic case, the phase shift between intensity and Doppler velocity (with positive values corresponding to redshifts) for propagating waves is $180^{\circ}$, while standing modes show a phase difference of $90^{\circ}$ \citep[\eg,][]{Deubner1974,Al+etal1998}. A phase shift of $90^{\circ}$ indicates that the intensity signal is delayed a quarter period with respect to the velocity signal. Figure \ref{fig:phase_V_I} illustrates the temporal evolution of the velocity and core intensity oscillations of the \HeI\ 10830 \AA\ filtered in the frequency band between 4.5 mHz and 8 mHz and averaged for all the spatial locations inside the umbra. The core intensity is defined as the intensity of the minimum from a second order polynomial fit to a spectral region of 0.23 \AA\ around the core of the line. Although the 4.5-8 mHz bandpass corresponds to propagating waves, according to the cutoff values observationally determined and plotted in Fig. \ref{fig:cutoffs}, during most of the temporal series illustrated in Fig. \ref{fig:phase_V_I} the intensity signal trails the velocity signal, as expected for standing waves. On the contrary, during some temporal spans they have opposite phases (for example, around $t=45$ min), as theory predicts for propagating waves. The spatially average phase shift over the whole temporal series between velocity and intensity is $108^{\circ}\pm14^{\circ}$, which departs from the $90^{\circ}$ predicted for standing waves. This is consistent with the phase shift measured between the velocity from two chromospheric lines (Fig. \ref{fig:dphase_CaIIHHe}), which proves that there is upward wave propagation at chromospheric heights. We interpret that slow magnetoacoustic waves are partially reflected at the transition region, forming a chromospheric standing wave \citep{Fleck+Deubner1989}, while the rest of the slow mode wave energy is propagating in the upward direction \citep{Jess+etal2012}.

\section{Conclusions}
\label{sect:conclusions}

We have evaluated the cutoff frequency in a sunspot umbra and its variation with the atmospheric height by analyzing the phase difference and amplification spectra between several pairs of spectral lines that probe different layers. Our measurements show that between the deep photosphere and high photosphere the cutoff frequency increases from 5 mHz to 6 mHz. At higher chromospheric values the cutoff is reduced to $\sim 3.1$ mHz. These results have been compared with the values obtained from the application of several analytical cutoff forms to the atmosphere inferred from the inversion of the photospheric lines and to a standard model of umbral stratification. This comparison reveals some significant differences at the photosphere.

Our study has strong implications on our understanding of wave propagation in magnetized structures and its applications to several topics of solar research. Sunspots are one of the main targets of local helioseismology. One of the key measurements of these techniques is the travel time of $p-$mode waves between two locations on the solar surface, generally filtered in frequency and phase speed. The travel-time shift is then defined as the difference between the travel time measured in a region of interest (for example, a sunspot) and that measured in a quiet Sun region. Recent works have found that the cutoff frequency is a main contributor to travel-time shift measurements \citep{Lindsey+etal2010, Schunker+etal2013, Felipe+etal2017b}. Due to the Wilson depression, the height of the upper turning point of the waves (that is, the height where the wave frequency equals the local cutoff frequency) is lowered. Thus, waves beneath sunspots travel a shorter path than in quiet Sun regions, and their travel time is reduced. Interpretation of the travel-time shifts can potentially benefit from simultaneous estimations of the atmospheric cutoff frequencies, like those carried out in this work. These data would complement the helioseismic measurements, and would help to define the solar layers sampled by the $p-$modes at each frequency.

Compressive waves are also one of the candidates to explain chromospheric heating \citep[\eg,][]{BelloGonzalez+etal2010b,Felipe+etal2011, Kanoh+etal2016, KrishnaPrasad+etal2017}. These studies usually quantify the available energy flux of the propagating waves at photospheric heights and, thus, a detailed characterization of the cutoff is fundamental to estimate which frequencies can actually contribute to the heating.

\begin{acknowledgements} 
We thank Dr. N. Bello Gonz\'alez for her support during the observations. Financial support from the Spanish Ministry of Economy and Competitivity through projects AYA2014-55078-P, AYA2014-60476-P and AYA2014-60833-P is gratefully acknowledged. C.K. was supported in part by grant DE 787/5-1 of the German Science Foundation (DFG). The 1.5-meter GREGOR solar telescope was built by a German consortium under the leadership of the Kiepenheuer-Institut f\"ur Sonnenphysik in Freiburg with the Leibniz-Institut f\"ur Astrophysik Potsdam, the Institut f\"ur Astrophysik
G\"ottingen, and the Max-Planck-Institut f\"ur Sonnensystemforschung in G\"ottingen as
partners, and with contributions by the Instituto de Astrof\'isica de Canarias and
the Astronomical Institute of the Academy of Sciences of the Czech Republic.
\end{acknowledgements}

\bibliographystyle{aa} % style aa.bst
\bibliography{biblio.bib}

\end{document}